# Symmetry and Hamiltonian structure of the scaling equation in isotropic turbulence

Zheng   Ran,   Shuqin Pan
(Shanghai Institute of Applied Mathematics and Mechanics,
Shanghai University, Shanghai 200072, P.R.China)

**Abstract**

The assumption of similarity and self-preservation, which permits an analytical determination of the energy decay in isotropic turbulence, has played an important role in the development of turbulence theory for more than half a century. Sedov (1944), who first found an ingenious way to obtain two equations from one. Nonethless, it appears that this problem has never been reinvestigated in depth subsequent to this earlier work. In the present paper, such an analysis is carried out, yielding a much more complete picture of self-preservation isotropic turbulence. Based on these exact solutions, some physically significant consequences of recent advances in the theory of self-preserved homogenous statistical solution of the Navier-Stokes equations are presented. New results could be obtained for the analysis on turbulence features, such as the scaling behavior, the spectrum, and also the large scale dynamics. The general energy spectra and their behavior in different wave number range are investigated. This paper focus on the scaling equation.

# 0. Introduction

Homogeneous isotropic turbulence is a kind of idealization for real turbulent motion, under the assumption that the motion is governed by a statistical law invariant for arbitrary translation (homogeneity), rotation or reflection (isotropy) of the coordinate system. This idealization was first introduced by Taylor (1935) and used to reduce the formidable complexity of statistical expression of turbulence and thus made the subject feasible for theoretical treatment. Up to now, a large amount of theoretical work has been devoted to this rather restricted kind of turbulence. However, turbulence observed either in nature or in laboratory has much more complicated structure. Although remarkable progress has been achieved so far in discovering various characteristics of turbulence, our understanding of the fundamental mechanism of turbulence is still partial and unsatisfactory.

The assumption of similarity and self-preservation, which permits an analytical determination of the energy decay in isotropic turbulence, has played an important role in the development of turbulence theory for more than half a century. In the traditional approach to search for similarity solutions for turbulence, the existence of a single length and velocity scale has been assumed, and then the conditions for the appearance of such solutions have been examined. Detailed research on the solutions of the Karman-Howarth equation was conducted by Sedov (1944), who showed that one could use the separability constraint to obtain the analytical solution of the Karman-Howarth equation. Sedov's solution could be expressed in terms of the confluent hypergeometric function. From the development of turbulence theory, we know that the research on decaying homogeneous isotropic turbulence is one of the most important and extensively explored topics. Despite all the efforts, a general theory describing the decay of turbulence based on the first principles has not yet been developed (Skrbek and Stalp, 2000). It seems that the theory of self-preservation in homogeneous turbulence has lots of interesting features which have not yet been fully understood and are worth of further study (see Speziale and Bernard,1992, p.665).

The present work will devote to the analysis on the scaling equation of sotropic turbulence. We will point out a new scaling solution set may exist if we adopt the Sedov method (1944). Hence, some new results could be obtained for revealing the features of turbulence. This paper focus on the two important issues:

[1] How the present work generalizes Sedov's?
[2] What's new in this generalization ?.

## 1. New scaling equation of isotropic turbulence

We will consider isotropic turbulence governed by the incompressible Navier-Stokes equations. The two-point double and triple longitudinal velocity correlation, denoted by $f(r,t)$ and $h(r,t)$ respectively, are defined in a standard way. For isotropic turbulence, they satisfy the Karman-Howarth equation

$$\frac{\partial}{\partial t}(bf) + 2b^{\frac{3}{2}}\left(\frac{\partial h}{\partial r} + \frac{4h}{r}\right) = 2nb\left(\frac{\partial^2 f}{\partial r^2} + \frac{4}{r}\frac{\partial f}{\partial r}\right) \qquad (1.1)$$

where $(r,t)$ is the spatial and time coordinates, $n$ is the kinematic viscosity, and $b = \overline{u^2}$ denotes the turbulence intensity. Let us suppose that the functions $f(r,t)$ and $h(r,t)$ preserve the same form as time increases with only the scale varying. Such functions will be termed as "self-preserving". Following von Karman and Howarth, we introduce the new variables

$$x = \frac{r}{l(t)} \qquad (1.2)$$

where $l = l(t)$ is a uniquely specified similarity length scale. For an isotropic turbulence to be self-preserving in the sense of von Karman and Howarth (1938) and Batchelor (1948),.For complete self-preserving isotropic turbulence, the Karman-Howarth equation takes the form

$$\frac{dh}{dx} + \frac{4}{x}h = -\frac{l}{2b^{\frac{3}{2}}}\frac{db}{dt}f + \frac{1}{2\sqrt{b}}\frac{dl}{dt}x\frac{df}{dx} + \frac{n}{\sqrt{bl}}\left(\frac{d^2f}{dx^2} + \frac{4}{x}\frac{df}{dx}\right) \qquad (1.3)$$

Sedov (1941) found an ingenious way to obtain two equations from one. The main results could be listed as following:
The turbulence system reduce to

$$\frac{d^2f}{dx^2} + \left(\frac{4}{x} + \frac{a_1}{2}x\right)\frac{df}{dx} + \frac{a_2}{2}f = 0 \qquad (1.4)$$

With boundary conditions $f(0) = 1$, $f(\infty) = 0$. and $a_1, a_2$ are constant coefficients.

The corresponding scaling (length and energy) equations read:

$$\frac{db}{dt} = -a_2\left(\frac{nb}{l^2}\right) \qquad (1.5)$$

$$\frac{d^2l}{dt^2} + \frac{(2a_1 + a_2)n}{2l^2}\frac{dl}{dt} - \frac{a_1 a_2 n^2}{2l^3} = 0. \qquad (1.6)$$

These are the scaling equations of isotropic turbulence .Substiuting leads the equation determining the third correlation coefficient

$$\frac{dh}{dx} + \frac{4}{x}h = \frac{p}{2}x\frac{df}{dx} - \frac{q}{2}f \qquad (1.7)$$

From the symmetry of $h$, we know that the expansion of $h$ in power of $x$ must start with the term of order $x^3$, consequenly, it folows from Eq.(1.7) that

$$q = 0 \qquad (1.8)$$

If we know the length scale, the substituting equation (1.5) will give the turbulence energy decay law. So, the departure point is how to deal with the equation (1.6). This is the main topic of this letter.

**2. Symmetry of the scaling equation**

Let
$$y(l) = \frac{dl}{dt} \tag{2.1}$$

Equ.(1.6) can recast into

$$yy'_l = -\frac{(2a_1 + a_2)n}{2l^2} \cdot y + \frac{a_1 a_2 n^2}{2l^3} \tag{2.2}$$

Symmetries and first integrals are two fundamental structures of ordinary differential equations (ODEs). Geometrically, it is natural to view an nth-order ODE as a surface in the $(n+2)$-dimensional space whose coordinates are given by the independent variable, the dependent variable and its derivates to order $n$, so that the solutions of the ODE are particular curves lying on this surface. From this point of view, symmetry represents a motion that moves each solution curve into solution curves; a first integral represent a quantity that is conserved along each solution curve. More precisely, symmetry is a one-parameter group of local transformations, acting on the coordinates involving the independent variable, the dependent variable and its derivatives to order $n-1$, that is constant on each solution. In this section, we show how to find admitted symmetries and first integrals of the turbulence scaling equation.

Based on the Lie symmetry analysis, equation (2.2) only admits a one-parameter Lie group of point transformation with infinitesimal

$$V = \frac{l}{\bar{a}} \frac{\partial}{\partial l} + \frac{y}{\bar{b}} \frac{\partial}{\partial y} \tag{2.3}$$

where $\bar{a}, \bar{b}$ are arbitrary constants satisfies

$$\frac{1}{\bar{a}} + \frac{1}{\bar{b}} = 0. \tag{2.4}$$

The corresponding characteristic equation is given by

$$\frac{\bar{a}}{l} dl = \frac{\bar{b}}{y} dy \tag{2.5}$$

The invariant

$$I_1 = \frac{y^{\bar{b}}}{l^{\bar{a}}} \tag{2.6}$$

This gives

$$y = \frac{dl}{dt} = [I_1]^{\frac{1}{\bar{b}}} \cdot l^{-1} \tag{2.7}$$

The solution in term of time could be

$$l^2 = 2[I_1]^{\frac{1}{\bar{b}}} \cdot (t + t_0) \tag{2.8}$$

It is well known that if an ODE admits a Lie group of transformations, then one can construct interesting special classes of solutions (invariant solutions) that correspond to invariant curves of the admitted by Lie group of transformations. For a first-order ODE, such invariant solutions can be determined algebraically. It is believed that this might be show some light on the nature of the power-law in scaling equation.

### 3. The solution of the first integrals

Based on the result of section 2, we are interested in the special solutions in the following form:

$$l(t) = l_0 (t + t_0)^{\frac{1}{2}} \tag{3.1}$$

where $l_0$ are parameters to be chosen.. Special attention would be paid to the derivation of $l_0$.

This equation can be recast into

$$l_0^4 - (2a_1 + a_2) n l_0^2 + 2a_1 a_2 n^2 = 0 \tag{3.2}$$

$$l_0^2 = \frac{n}{2}\left\{(2a_1 + a_2) \pm \sqrt{(2a_1 - a_2)^2}\right\} \tag{3.3}$$

Here, we would like to introduce some notations for the convenience.

If

$$l_0^2 = \frac{n}{2}\left\{(2a_1 + a_2) + \sqrt{(2a_1 - a_2)^2}\right\} \tag{3.4}$$

we denote its by P_mode.

If

$$l_0^2 = \frac{n}{2}\left\{(2a_1 + a_2) - \sqrt{(2a_1 - a_2)^2}\right\} \tag{3.5}$$

we denote its by N_mode.

But the final results will depend on the values of $s$. The definition of $s$ is

$$s \equiv \frac{a_2}{2a_1} \tag{3.6}$$

The details of the solutions are listed in Table.1.

### Table.1 Values of $l_0^2$

|  | **P_mode** | **N_mode** |
| --- | --- | --- |
| $0 < s \leq 1$ | $2a_1 n$ | $a_2 n$ |
| $s > 1$ | $a_2 n$ | $2a_1 n$ |

### 4. Hamilton structure of the scaling equation

Recently (Gladwn, et al., 2009), it is pointed out the existence of a remarkable nonlocal transformation between the damped harmonic oscillator and a modified Emden-type nonlinear oscillator equation with linear force, which preserves the form of the time independent integral, conservative Hamiltonian, and the equation of motion. Generalizing this transformation we prove the existence of nontandard conservative Hamiltonian structure for a general class of damped nonlinear oscillators including Lienard-type systems. Further, using the above Hamiltonian structure for a specific example, namely, the generalized modified Emden equation, the general solution is obtained through appropriated canonical transformations.

In this section, we point out the canonical equations of motion for turbuelnce scaling equaton now read

$$\begin{cases} \dot{U} = \dfrac{\partial H}{\partial P} = s_1 n_1 P^{n_1-1} U^{m_1} \\ \dot{P} = -\dfrac{\partial H}{\partial U} = -(s_1 m_1 P^{n_1} U^{m_1-1} + 2h_1 U) \end{cases} \quad (4.1)$$

Where

$$H = s_1 P^{n_1} U^{m_1} + h_1 U^2$$

$$s_1 = \frac{1}{r-2}\left(\frac{r-1}{2}\right)^{\frac{r-2}{r-1}}, \quad n_1 = \frac{r-2}{2(r-1)}, \quad m_1 = \frac{3(r-2)}{2(r-1)},$$

$$h_1 = \frac{1-r}{8r}\left[1+\frac{1}{s}\right]a_2 n$$

$$r = \frac{5}{3}+n, \quad s = \frac{2}{3}+n, \quad n = 0,1,2,\mathbf{L},N$$

Figure 1. show the different phase portraits of equation (4.1) with different values of $n$ for $a_2 n = 1$. This might be help to understanding the micro-structure of isotropic turbulence.

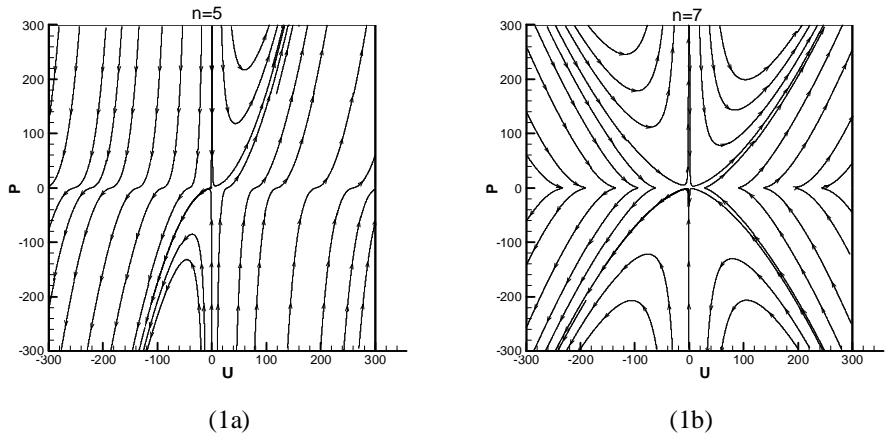

(1a)          (1b)

**Figure.1 Phase portraits of turbulence scaling equation**

**5. Conclusion**

Based on the new scaling eqution of isotropic turbulence, the associated symmetry and its Hamiltonian structrue are presented in this letter. A richer mathematical structure in this letter will help to the micro-structure of isotropic turblence. Such a conserved Hamiltonian description also leads one to further investigation the instability in isotropic turbulence.

**Acknowledgements:**

The work was supported by the National Natural Science Foundation of China (Grant Nos.90816013, 10572083),